\documentclass[fleqn,usenatbib]{mnras}

\usepackage{newtxtext,newtxmath}

\usepackage[T1]{fontenc}

\DeclareRobustCommand{\VAN}[3]{#2}
\let\VANthebibliography\thebibliography
\def\thebibliography{\DeclareRobustCommand{\VAN}[3]{##3}\VANthebibliography}

\usepackage{graphicx}	
\usepackage{amsmath}	

\newcommand{\xHIv}{{\left< x_{\rm HI}\right>_{\rm v}}}
\newcommand{\Lya}{Ly$\alpha $ }
\newcommand{\CHIMP}{~h^{-1}{\rm~cMpc}}
\newcommand{\Msun}{{\rm M}_\odot}

\title[Damping wings]{The Characteristic Shape of Damping Wings During Reionization}

\author[Huanqing Chen]{
Huanqing Chen$^{1}$\thanks{E-mail: hqchen@cita.utoronto.ca}
\\
$^{1}$Canadian Institute for Theoretical Astrophysics, University of Toronto,60 St George St, Toronto, ON M5R 2M8, Canada
}

\date{Accepted XXX. Received YYY; in original form ZZZ}

\pubyear{2015}

\begin{document}
\label{firstpage}
\pagerange{\pageref{firstpage}--\pageref{lastpage}}
\maketitle

\begin{abstract}
Spectroscopic analysis of Ly$\alpha$ damping wings of bright sources at $z>6$ is a promising way to measure the reionization history of the universe. However, the theoretical interpretation of the damping wings is challenging due to the inhomogeneous nature of the reionization process and the proximity effect of bright sources. In this Letter, we analyze the damping wings arising from the neutral patches in the radiative transfer cosmological simulation suite Cosmic Reionization on Computers (CROC). We find that the damping wing profile remains a tight function of volume-weighted neutral fraction $\xHIv$, especially when $\xHIv>0.5$, despite the patchy nature of reionization and the proximity effect. 
This small scatter indicates that with a well-measured damping wing profile, we could constrain the volume-weighted neutral fraction as precise as $\Delta \xHIv\lesssim 0.1$ in the first half of reionization. 
\end{abstract}




\section{Introduction}
The epoch of reionization (EoR) brought about a major change to the global properties of the intergalactic medium (IGM) within the first billion years of the universe. Thanks to the numerous data obtained by JWST, we are now in an excellent position to understand this frontier in astrophysics. 

One of the fundamental questions surrounding the EoR concerns the timing and duration of reionization, which is not yet well-constrained.
Several methods have been employed to measure reionization, each with its own unique strengths and limitations.
One of the earliest constraints came from cosmic microwave background experiments that utilized the Thompson scattering effects of free electrons. For example, \citet[][]{planck2020} has constrained the midpoint of reionization redshift to be $z_{\rm mid}=7.82 \pm 0.71 $ \citep[see also][]{spt2012,act2011}. However, accurately measuring the entire history of reionization from start to end using Thompson scattering effect on CMB alone is challenging due to its integrated nature. Another powerful method to characterize the full cosmic dawn and reionization history is through the 21cm line emission from neutral hydrogen. However, at such long wavelengths, the foreground is orders of magnitude brighter than the signals, making data reduction notoriously difficult \citep[e.g.,][]{kern2021,mertens2020}. 

Another alternative of measuring the entire reionization history is to use the \Lya  absorption in front of bright sources at different redshift during the EoR \citep[e.g.,][]{miralda-escude1998, totani2016, greig2017, davies2018, greig2019, greig2022, gnedin2004, mason2018, mason2020}. As a strong resonant line, \Lya is sensitive to any trace of neutral hydrogen, allowing us to detect the very end of reionization (neutral fraction $\lesssim 10^{-4}$) \citep[e.g.,][]{bosman2018,zhu2021,bosman2022}. Moreover, when there are neutral patches left in the IGM, the \Lya absorption line displace a large damping wing, reaching thousands of km/s in the spectrum where the flux is suppressed. \citet{miralda-escude1998} shows that assuming a uniform reionization model, the damping wings have characteristic shape and can be used in constrain neutral fraction for bright background sources like gamma-ray bursts (GRBs). 

However, reionization is a patchy process. The neutral fraction does not drop  uniformly everywhere. Rather, some regions become highly ionized first, while other regions, shielded from ionizing sources, remain neutral until much later. In fact, many semi-numerical codes based on excursion-set formalism \citep[e.g.,][]{furlanetto2004, mesinger2007} treat every point in the universe as either neutral or ionized, therefore the term neutral fraction is meaningful only when averaged over certain volume or mass. In the literature of reionization, the term neutral fraction most commonly refers to the volume-weighted neutral fraction over the entire universe $\xHIv$ \citep[e.g.,][]{zhu2021}.

Given the patchy nature of reionization, one natural question is whether the variance of the damping wing profile is too large to differentiate universes with different $\xHIv$, or if it is small enough that the characteristic shape still holds.
Another complication arises from the fact that bright sources, such as quasars, which provide high-resolution spectra for analysis, emit a large amount of ionizing radiation themselves. This radiation can alter the local morphology of reionization.
Many semi-numerical methods can create a map of ionized bubbles created from typical galaxies \citep[e.g.,][]{furlanetto2004, mesinger2007, mesinger2011, zahn2011}, but unusual bright sources like quasars are not modeled. Does the removal of neutral patches close to bright sources like quasars significantly change the shape of the damping wing?

In this Letter, we use a radiative transfer cosmological simulation suite Cosmic Reionization on Computers (CROC) to address the above questions. Such a study is timely as more and more bright sources at $z>6$  are spectroscopically followed-up and available to be used in constraining reionization history. The letter does not intend to describe the full process of extracting neutral fraction from data, but serves to estimate the optimal precision of neutral fraction measurement achievable using \Lya damping wing.

\begin{figure*}
    \centering
    \includegraphics[width=0.33\textwidth]{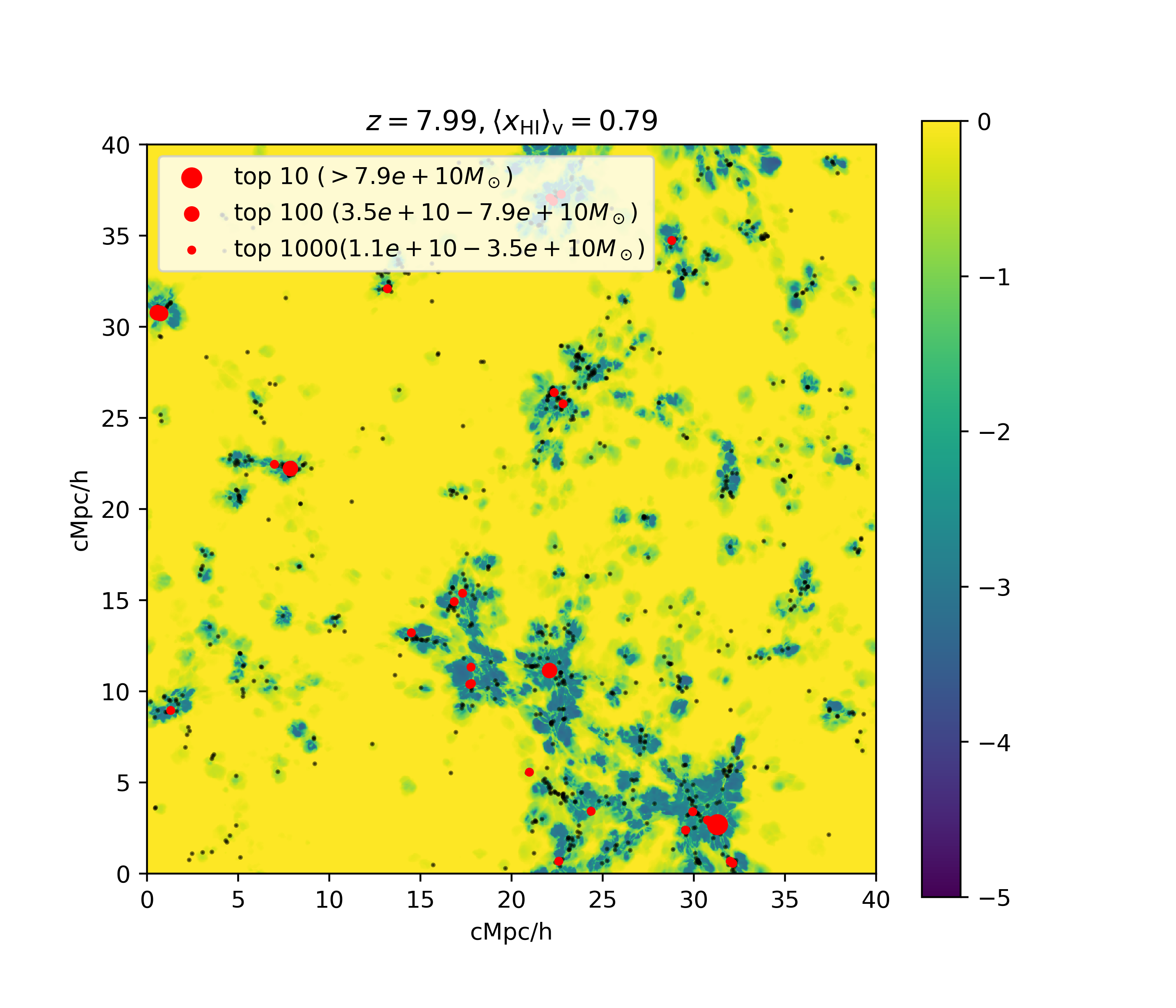}
    \includegraphics[width=0.33\textwidth]{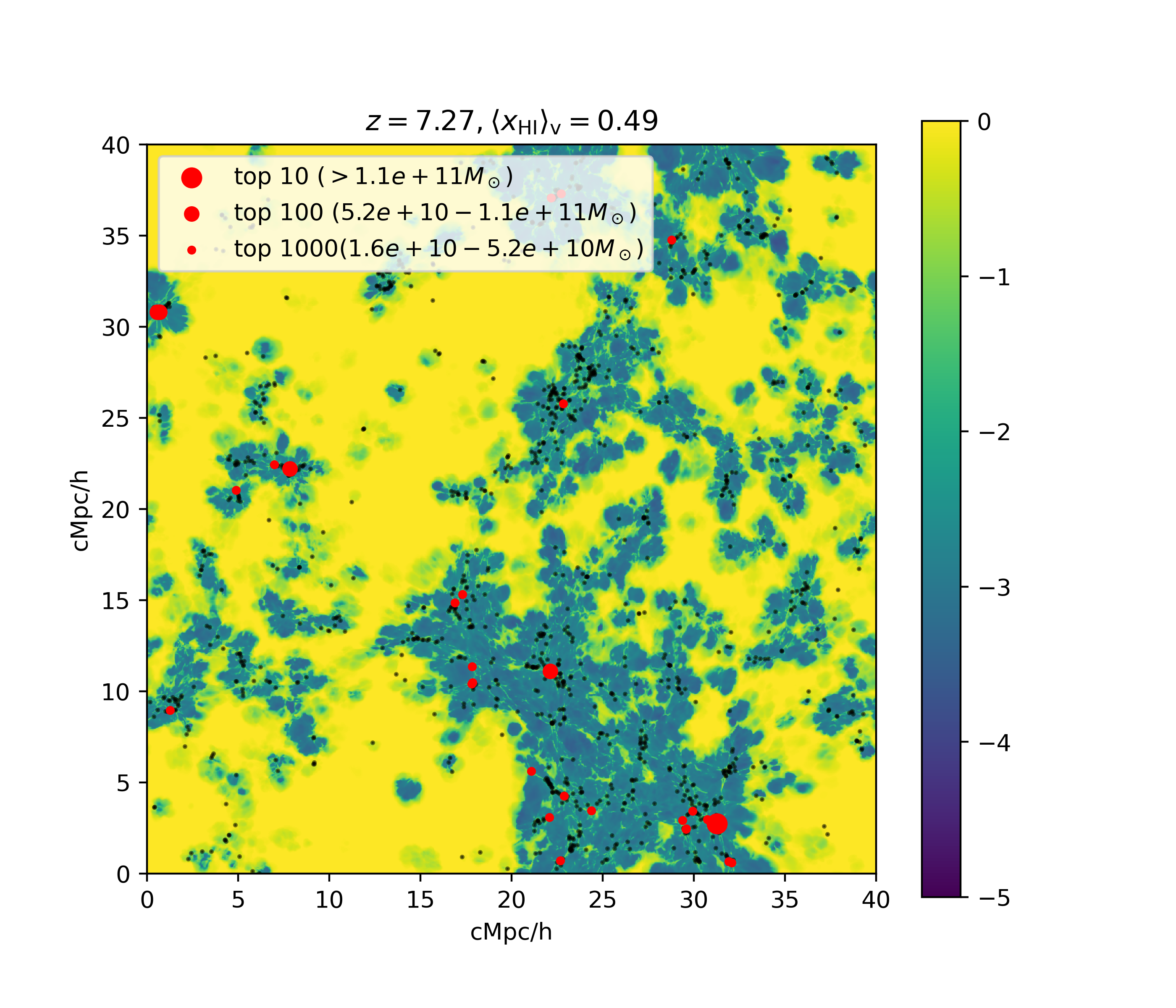}
    \includegraphics[width=0.33\textwidth]{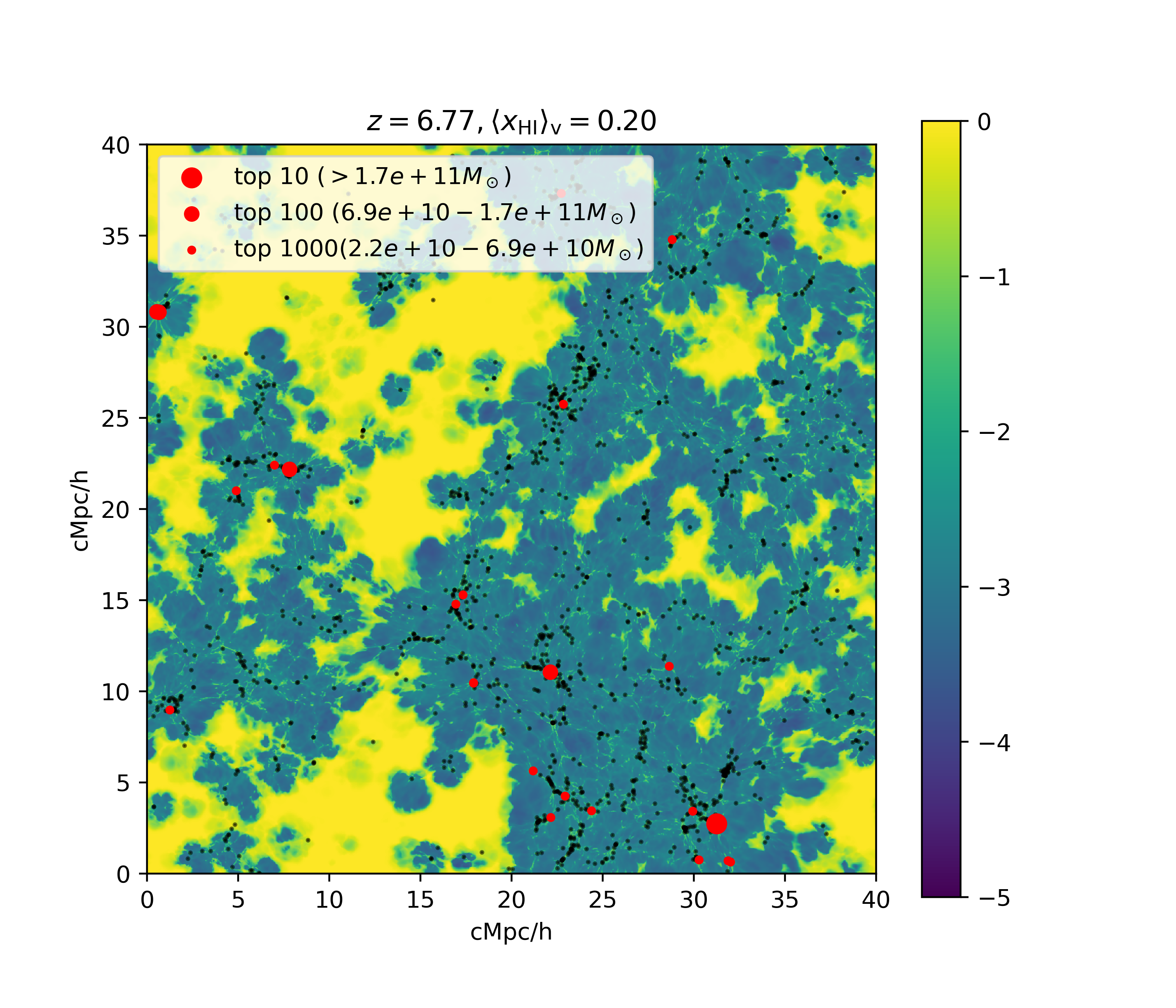}
    \caption{Neutral fraction map in the simulation (CROC B40F), overlaid with halos of different masses. Red dots represent top 1000 massive halos and black dots are all halos with mass larger than $10^9 \Msun$. Each panel show a slice of thickness of $2$ cMpc. The colorbar shows the neutral fraction of each cell in logarithmic scale.}
    \label{fig:data}
\end{figure*}

\begin{figure*}
    \centering\includegraphics[width=\textwidth]{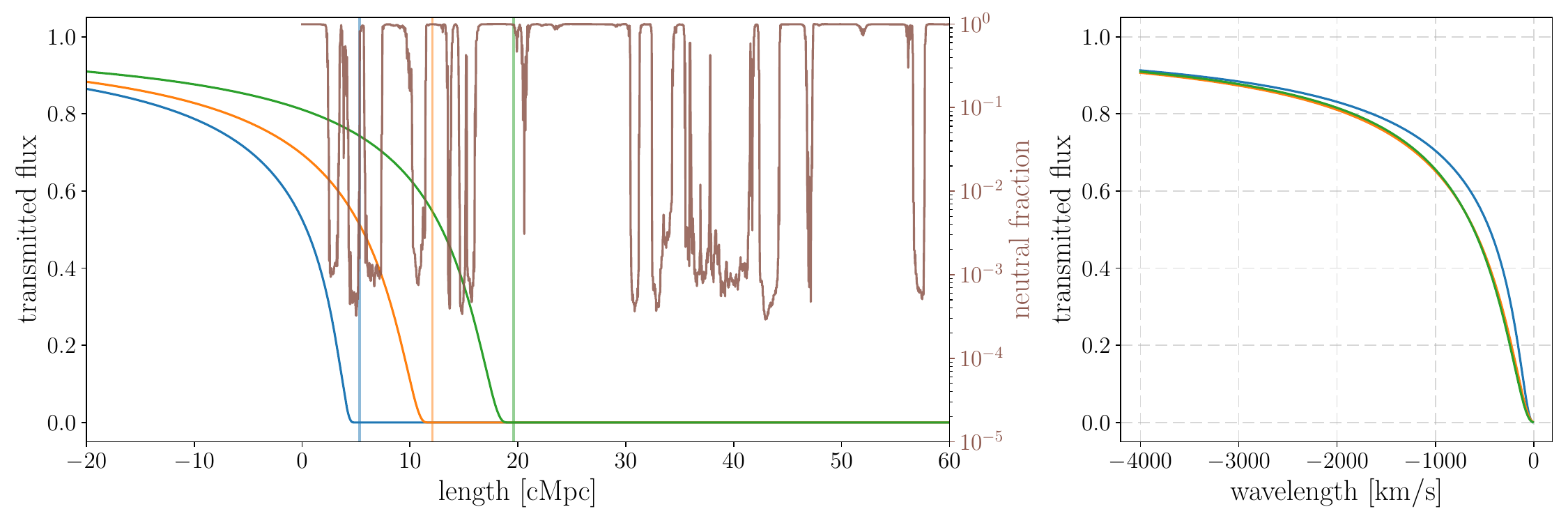}
    \caption{Procedure to produce damping wing profiles for snapshot $\xHIv=0.5$. Left panel: the brown line shows the neutral fraction vs distance from a massive halo. Faint vertical lines show three different random positions within which we cut out inner neutral gas, and the solid lines show the shape of damping wing created by all neutral gas behind it until $200 ~ {\rm cMpc}/h$. Right panel: damping wing profiles in the left panel but moved to the same starting point.}
    \label{fig:cut_wing} 
\end{figure*}
\section{Simulation}
We use CROC simulations\footnote{The cosmological parameters used in CROC are: $\Omega_b=0.0479, \Omega_M=3036, \Omega_\Lambda=0.6964, h=0.6814, n_s=0.9675, \sigma_8=0.8285, k_{\rm pivot}=0.029$.} \citep{gnedin2014} to study the damping wings arising from patchily ionized IGM. The CROC project uses the Adaptive Refinement Tree (ART) code  \citep{kravtsov1999, kravtsov2002, rudd2008} to reach high spatial resolution (base grid length $=39\  h^{-1}~\rm~ckpc$, peak resolution $\sim 100$ pc in physical units).
CROC simulations include relevant physics such as gas cooling, heating, star formation, stellar feedback and on-the-fly radiative transfer \citep{gnedin2001}. The main ionization sources in the simulations are star particles which are formed in dense gas in galaxies.

In this project, we primarily use the uniform-grid data in one of the $40 \CHIMP $ runs (CROC B40F) alongside with Rockstar \citep{behroozi2013} halo catalogs to locate dark matter halos. The uniform-grid data contain gas properties of neutral fraction, density, temperature in each base grid cell. 
They are saved frequently (with increments in expansion factor $\Delta a=0.001$) so that we can sample a large range of $\xHIv$ and study the entire reionization process.
In Figure \ref{fig:data}, we show the neutral fraction map at three different redshifts overlaid with halos of different masses.

\section{Results}\label{sec:results}

To simulate the damping wing profiles, we first draw skewers (sightlines) starting from massive halos.
We locate halos from Rockstar halo catalogues in the uniform-grid box. At each redshift, we select the $100$ most massive halos and draw $10$ skewers of length $200 ~ {\rm cMpc}/h$ uniformly distributed in a 3D sphere. In the left panel of Figure \ref{fig:cut_wing}, the brown line shows the neutral fraction along one example skewer drawn from a snapshot where the volume-weighted neutral fraction is $\xHIv=0.5$.
To study the universe with different neutral fractions $\xHIv$, we use skewers drawn at different redshifts of the same simulation run. When calculating \Lya absorption, we keep the neutral fraction and temperature of each cell unchanged while scale the physical length and density to a certain redshift $z_{t}$ by $a$ and $a^{-3}$, where $a$ is the expansion factor, respectively. The results shown in this paper are calculated for $z_{t}=6.54$.

An unusually bright source like a quasar could push the I-front farther away. To mimic such an extra ionizing effect, when calculating the damping wing, we first draw a random number from a uniform distribution between [0, 40] ${\rm cMpc}/h$ and remove all neutral gas within this distance. This procedure aims to examine the maximum variance of the damping wing shape. Then we convolve the Voigt profiles from the rest of the neutral cells ($x_{\rm HI}>0.5$) along the skewer. In the left panel of Figure \ref{fig:cut_wing}, we show this procedure in a skewer drawn from the box with $\xHIv=0.5$: the faint blue, orange and green vertical lines show three random positions within which we remove all neutral gas, and the solid profiles are the damping wing arising from the remaining neutral gas, integrated until $200 ~ {\rm cMpc}/h$. We find that despite the length of the first neutral patch are different, after convolution with all neutral patches behind it, the shapes of the profiles are very similar. This is more evident in the right panel, where we compare these profiles after aligning them at the starting position (the first point where transmission drops to zero).

\begin{figure}
    \centering\includegraphics[width=0.45\textwidth]{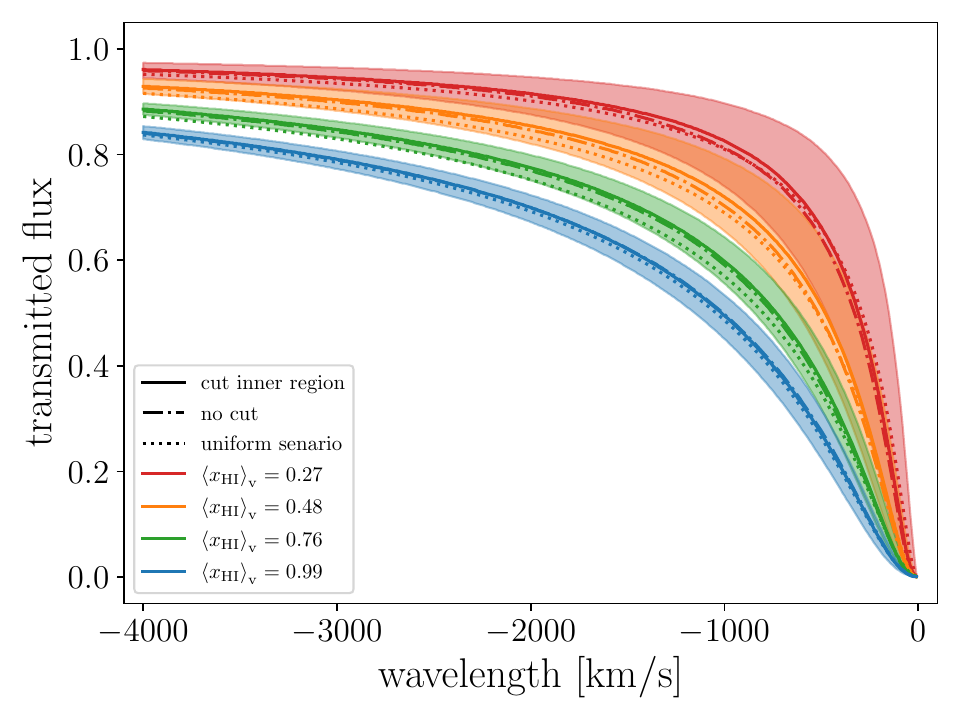}
    \caption{Damping wing profiles at different global volume-weighted neutral fraction $\xHIv$. Lines show the median while bands show $68\%$ scatter. The solid lines show profiles created with a randomly excised inner region as described in Sec. \ref{sec:results}. The dash-dotted lines show profiles which  we do not excise a random inner region and the dotted lines show the cases of uniform reionization.}
    \label{fig:wing_vs_xHIv} 
\end{figure}

In Figure \ref{fig:wing_vs_xHIv}, we plot the median of the aligned damping wing profiles in snapshots of different $\xHIv$ using solid lines, with each colored band showing the $68\%$ scatter. 
For $\xHIv >=0.5$, the scatter of the wing profile is very small despite the patchy nature of reionization, and profiles with $\Delta \xHIv =0.25$ are clearly separated. We also compare the damping wing profiles with the ones that created without randomly cutting the inner region (dash-dotted lines). 
If the inner neutral regions are not excised, the median damping wing is slightly stronger, but well within the scatter. The scatter of the no-cut case is almost identical to the previous case and thus not shown. We also calculate the damping wings assuming a uniform density and uniform reionization scenario (every cell has the same neutral fraction $x_{\rm HI}=\xHIv$ and every cell contributes to the damping wing), and the results are shown as dotted lines. Compared with the patchy ionization scenario with inner region excised, the damping wing is in general stronger, especially for $0.5\lesssim \xHIv\lesssim 0.75$, but the differences are still small compared to the scatter.

\section{Discussion}
\subsection{Cosmic variance}

\begin{figure}
    \centering\includegraphics[width=0.45\textwidth]{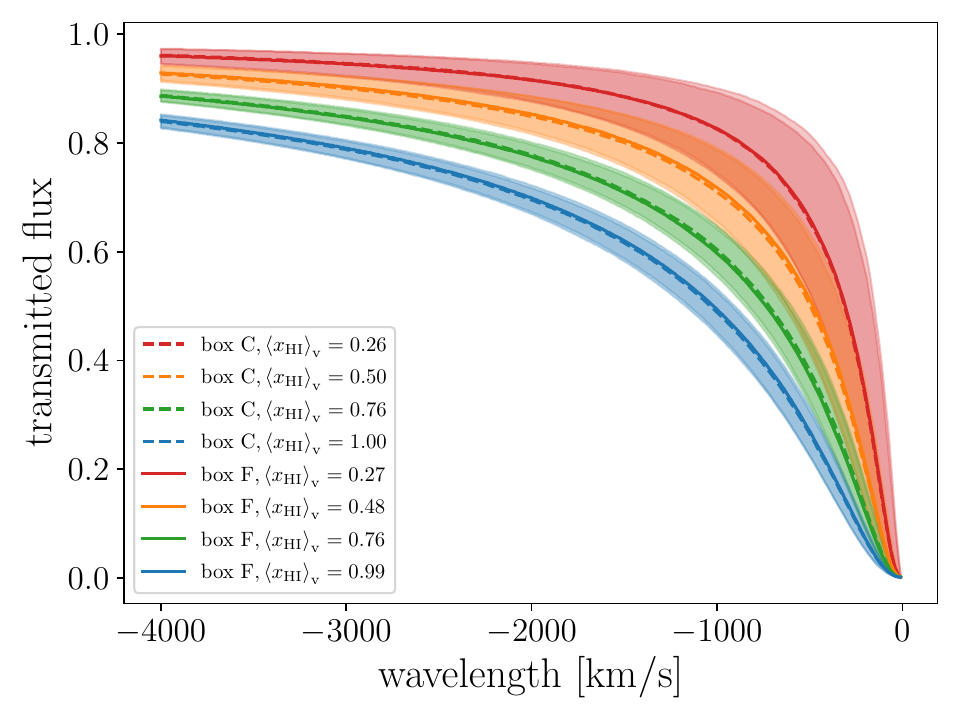}
    \caption{Comparing damping wing shapes in two different CROC boxes (B40C and B40F).}
    \label{fig:boxC_boxF} 
\end{figure}

Due to the small size ($40 \CHIMP$) of the simulation box, one might question whether the small scatter shown in the last section still holds when considering cosmic variance. To investigate this, we repeat the procedure in another box (CROC B40C). Both simulations have the same physics but different initial conditions (``DC modes''). As a result, B40F reionized the latest (reionization midpoint $z_{\rm mid}=7.4$) while B40C reionized earliest ($z_{\rm mid}=8.2$) in all the six $40 \CHIMP$ CROC realizations. Therefore, the density environments and halo distribution in these two box should differ maximally in all six realizations, and comparing damping wings in these two boxes helps us understand the stochasticity due to cosmic variance. In Figure \ref{fig:boxC_boxF}, we compare the damping wings in box B40C with B40F of the previous section. We find that the mean and the scatter are almost identical, suggesting that the damping wings indeed have a characteristic shape as a function of $\xHIv$.

\subsection{Practical use}
Our simulations show that for a mostly neutral universe ($\xHIv$ >
0.5), the scatter in damping wing profiles is small enough to distin-
guish between $\Delta \xHIv \approx 0.1$. However, measuring the entire damping wing profile is complicated in practice. In this subsection we briefly discuss the prospects of using damping wing to constrain $\xHIv$.

\citet{miralda-escude1998} originally proposes GRB afterglows as the best candidates for measuring $\xHIv$ with damping wings. Compared with galaxies or quasars, GRBs have many advantages. They are thought to be produced in normal galaxies and thus live in less biased environments \citep{woosley1993}. The number of integrated ionizing photons they contribute is also very small and unlikely to enlarge the local ionized bubble. In addition, they are intrinsically bright to be spectroscopically followed-up. One challenge of using GRB afterglows is how to model the Damped \Lya absorbers (DLAs) in the host galaxies. \citet[][]{lidz2021} shows that using the empirical distribution from current GRB afterglow spectra, one could model the local DLA distribution and marginalize this nuance parameter. Although \citet[][]{lidz2021} does not consider the scatter of damping wing profiles, the small scatter we find in CROC simulations supports their forecast that with $\gtrsim 20$ GRB afterglows with spectra resolution $R\gtrsim 3000$ and signal-to-noise ratio (SNR) $\gtrsim 20$, one could reach a precision close to $15\%$ in the first half of reionization.

Quasars are the kind of sources we can obtain the highest resolution spectra at $z>6$. The current highest quality sample of $z>6$ quasar spectra have SNR$\gtrsim 50$ and $R\gtrsim 10000$ \citep{DOdorico2023}. Thanks to their high luminosity, the residual neutral fraction in their proximity zone is small enough to allow significant flux on the blue side of \Lya line. Such flux offers extra information about the shape of the damping wing. The challenge of using quasars is that by $z\approx 7$, a quasar may have enlarged the local bubble significantly. Due to the decrease in quasar radiation with distance, the transmitted \Lya flux also decreases. This reduction in flux compromises the constraining power on the starting point of the damping wing, which is crucial for anchoring the shape of the damping wing. In the ideal case, we may catch a quasar in its bright phase, where the integrated ionizing photons emitted by the quasar is still small while the instantaneous luminosity is high enough to create a highly transparent proximity zone. This would allow us to observe details of the \Lya forest and measure flux close to the starting point of the damping wing, providing greater constraining power for the shape of the entire damping wing. In addition, similar to the GRB afterglow case, we need to develop a better understanding of how to model the intrinsic quasar continua. Since the scatter in damping wing profile is $< 10\%$ at wavelength $<-1000$ km/s from the starting point of the damping wing, it is ideal to have an accuracy in continuum recovery better than $10 \%$ across the quasar \Lya emission line (from $\lesssim -4000$ km/s to where no transmitted flux presents).

With the successful operation of JWST, we now have the capability to measure spectra from Lyman Break Galaxies (LBGs) or Lyman-alpha Emitters (LAEs) \citep[e.g.,][]{bolan2022}. While these sources are more numerous than quasars, their low luminosity limits the achievable spectral resolution. As a result, information about the \Lya damping wing is mainly contained in the equivalent width (EW) measurements. However, if we can combine the information of both the LBG/LAE positions and their EWs, it would be promising to constrain the neutral fraction by considering both the damping wing strength and the size of ionized bubbles \citep[e.g.,][]{mason2020,tang2023}. This avenue will be explored in future work.

\section{Conclusions}
In this paper, we analyze the damping wings arisen from the partially ionized IGM in a self-consistent radiative transfer cosmological simulation suite CROC. We find that when the volume-weighted neutral fraction $\left< x_{\rm HI} \right> > 0.5$, the shape of the damping wing has a characteristic shape with small scatter ($\lesssim 10\%$). This scatter remains small even after an unusually bright source (such as a quasar) erodes a significant amount of neutral gas around it. This is because the damping wing arises from the collective, convoluted Voigt profiles along a large distance (hundreds of comoving Mpc). We also calculate the damping wing profiles in a uniform reionization case, and we find that it lies within the $68\%$ scatter.

The small scatter in the damping wing profiles indicates that we can expect an accuracy of $\Delta \xHIv \approx 0.1$ if we could measure the damping wing profile precisely. In reality, there are several complications, notably how to model the intrinsic source spectra and the \Lya absorption within the ionized bubble. The profiles we find suggest that in order to achieve the best constraints in neutral fraction, we should aim for an accuracy of continuum fitting better than $10\%$ across the \Lya emission line of the source (from $\lesssim -4000$ km/s to where no transmitted flux presents). For a very bright source such as a quasar, the complication of \Lya absorption inside the ionized bubble could potentially be mitigated by properly modeling the large-scale structure, which we plan to explore in the future.

\section*{Acknowledgements}
HC thanks the support by the Natural Sciences and Engineering Research Council of Canada (NSERC), funding reference \#DIS-2022-568580.

\section*{Data Availability}

 The data underlying this article will be shared on reasonable request
to the author.



\bibliographystyle{mnras}
\bibliography{ms} 

\begin{thebibliography}{}
\makeatletter
\relax
\def\mn@urlcharsother{\let\do\@makeother \do\$\do\&\do\#\do\^\do\_\do\%\do\~}
\def\mn@doi{\begingroup\mn@urlcharsother \@ifnextchar [ {\mn@doi@}
  {\mn@doi@[]}}
\def\mn@doi@[#1]#2{\def\@tempa{#1}\ifx\@tempa\@empty \href
  {http://dx.doi.org/#2} {doi:#2}\else \href {http://dx.doi.org/#2} {#1}\fi
  \endgroup}
\def\mn@eprint#1#2{\mn@eprint@#1:#2::\@nil}
\def\mn@eprint@arXiv#1{\href {http://arxiv.org/abs/#1} {{\tt arXiv:#1}}}
\def\mn@eprint@dblp#1{\href {http://dblp.uni-trier.de/rec/bibtex/#1.xml}
  {dblp:#1}}
\def\mn@eprint@#1:#2:#3:#4\@nil{\def\@tempa {#1}\def\@tempb {#2}\def\@tempc
  {#3}\ifx \@tempc \@empty \let \@tempc \@tempb \let \@tempb \@tempa \fi \ifx
  \@tempb \@empty \def\@tempb {arXiv}\fi \@ifundefined
  {mn@eprint@\@tempb}{\@tempb:\@tempc}{\expandafter \expandafter \csname
  mn@eprint@\@tempb\endcsname \expandafter{\@tempc}}}

\bibitem[\protect\citeauthoryear{{Behroozi}, {Wechsler}  \& {Wu}}{{Behroozi}
  et~al.}{2013}]{behroozi2013}
{Behroozi} P.~S.,  {Wechsler} R.~H.,   {Wu} H.-Y.,  2013, \mn@doi [\apj]
  {10.1088/0004-637X/762/2/109}, \href
  {https://ui.adsabs.harvard.edu/abs/2013ApJ...762..109B} {762, 109}

\bibitem[\protect\citeauthoryear{{Bolan} et~al.,}{{Bolan}
  et~al.}{2022}]{bolan2022}
{Bolan} P.,  et~al., 2022, \mn@doi [\mnras] {10.1093/mnras/stac1963}, \href
  {https://ui.adsabs.harvard.edu/abs/2022MNRAS.517.3263B} {517, 3263}

\bibitem[\protect\citeauthoryear{{Bosman}, {Fan}, {Jiang}, {Reed}, {Matsuoka},
  {Becker}  \& {Haehnelt}}{{Bosman} et~al.}{2018}]{bosman2018}
{Bosman} S. E.~I.,  {Fan} X.,  {Jiang} L.,  {Reed} S.,  {Matsuoka} Y.,
  {Becker} G.,   {Haehnelt} M.,  2018, \mn@doi [\mnras]
  {10.1093/mnras/sty1344}, \href
  {https://ui.adsabs.harvard.edu/abs/2018MNRAS.479.1055B} {479, 1055}

\bibitem[\protect\citeauthoryear{{Bosman} et~al.,}{{Bosman}
  et~al.}{2022}]{bosman2022}
{Bosman} S. E.~I.,  et~al., 2022, \mn@doi [\mnras] {10.1093/mnras/stac1046},
  \href {https://ui.adsabs.harvard.edu/abs/2022MNRAS.514...55B} {514, 55}

\bibitem[\protect\citeauthoryear{{D'Odorico} et~al.,}{{D'Odorico}
  et~al.}{2023}]{DOdorico2023}
{D'Odorico} V.,  et~al., 2023, \mn@doi [\mnras] {10.1093/mnras/stad1468}, \href
  {https://ui.adsabs.harvard.edu/abs/2023MNRAS.523.1399D} {523, 1399}

\bibitem[\protect\citeauthoryear{{Davies} et~al.,}{{Davies}
  et~al.}{2018}]{davies2018}
{Davies} F.~B.,  et~al., 2018, \mn@doi [\apj] {10.3847/1538-4357/aad6dc}, \href
  {https://ui.adsabs.harvard.edu/abs/2018ApJ...864..142D} {864, 142}

\bibitem[\protect\citeauthoryear{{Dunkley} et~al.,}{{Dunkley}
  et~al.}{2011}]{act2011}
{Dunkley} J.,  et~al., 2011, \mn@doi [\apj] {10.1088/0004-637X/739/1/52}, \href
  {https://ui.adsabs.harvard.edu/abs/2011ApJ...739...52D} {739, 52}

\bibitem[\protect\citeauthoryear{{Furlanetto}, {Zaldarriaga}  \&
  {Hernquist}}{{Furlanetto} et~al.}{2004}]{furlanetto2004}
{Furlanetto} S.~R.,  {Zaldarriaga} M.,   {Hernquist} L.,  2004, \mn@doi [\apj]
  {10.1086/423025}, \href
  {https://ui.adsabs.harvard.edu/abs/2004ApJ...613....1F} {613, 1}

\bibitem[\protect\citeauthoryear{{Gnedin}}{{Gnedin}}{2014}]{gnedin2014}
{Gnedin} N.~Y.,  2014, \mn@doi [\apj] {10.1088/0004-637X/793/1/29}, \href
  {https://ui.adsabs.harvard.edu/abs/2014ApJ...793...29G} {793, 29}

\bibitem[\protect\citeauthoryear{{Gnedin} \& {Abel}}{{Gnedin} \&
  {Abel}}{2001}]{gnedin2001}
{Gnedin} N.~Y.,  {Abel} T.,  2001, \mn@doi [\na]
  {10.1016/S1384-1076(01)00068-9}, \href
  {https://ui.adsabs.harvard.edu/abs/2001NewA....6..437G} {6, 437}

\bibitem[\protect\citeauthoryear{{Gnedin} \& {Prada}}{{Gnedin} \&
  {Prada}}{2004}]{gnedin2004}
{Gnedin} N.~Y.,  {Prada} F.,  2004, \mn@doi [\apjl] {10.1086/422390}, \href
  {https://ui.adsabs.harvard.edu/abs/2004ApJ...608L..77G} {608, L77}

\bibitem[\protect\citeauthoryear{{Greig}, {Mesinger}, {Haiman}  \&
  {Simcoe}}{{Greig} et~al.}{2017}]{greig2017}
{Greig} B.,  {Mesinger} A.,  {Haiman} Z.,   {Simcoe} R.~A.,  2017, \mn@doi
  [\mnras] {10.1093/mnras/stw3351}, \href
  {https://ui.adsabs.harvard.edu/abs/2017MNRAS.466.4239G} {466, 4239}

\bibitem[\protect\citeauthoryear{{Greig}, {Mesinger}  \& {Ba{\~n}ados}}{{Greig}
  et~al.}{2019}]{greig2019}
{Greig} B.,  {Mesinger} A.,   {Ba{\~n}ados} E.,  2019, \mn@doi [\mnras]
  {10.1093/mnras/stz230}, \href
  {https://ui.adsabs.harvard.edu/abs/2019MNRAS.484.5094G} {484, 5094}

\bibitem[\protect\citeauthoryear{{Greig}, {Mesinger}, {Davies}, {Wang}, {Yang}
  \& {Hennawi}}{{Greig} et~al.}{2022}]{greig2022}
{Greig} B.,  {Mesinger} A.,  {Davies} F.~B.,  {Wang} F.,  {Yang} J.,
  {Hennawi} J.~F.,  2022, \mn@doi [\mnras] {10.1093/mnras/stac825}, \href
  {https://ui.adsabs.harvard.edu/abs/2022MNRAS.512.5390G} {512, 5390}

\bibitem[\protect\citeauthoryear{{Kern} \& {Liu}}{{Kern} \&
  {Liu}}{2021}]{kern2021}
{Kern} N.~S.,  {Liu} A.,  2021, \mn@doi [\mnras] {10.1093/mnras/staa3736},
  \href {https://ui.adsabs.harvard.edu/abs/2021MNRAS.501.1463K} {501, 1463}

\bibitem[\protect\citeauthoryear{{Kravtsov}}{{Kravtsov}}{1999}]{kravtsov1999}
{Kravtsov} A.~V.,  1999, PhD thesis, NEW MEXICO STATE UNIVERSITY

\bibitem[\protect\citeauthoryear{{Kravtsov}, {Klypin}  \& {Hoffman}}{{Kravtsov}
  et~al.}{2002}]{kravtsov2002}
{Kravtsov} A.~V.,  {Klypin} A.,   {Hoffman} Y.,  2002, \mn@doi [\apj]
  {10.1086/340046}, \href
  {https://ui.adsabs.harvard.edu/abs/2002ApJ...571..563K} {571, 563}

\bibitem[\protect\citeauthoryear{{Lidz}, {Chang}, {Mas-Ribas}  \& {Sun}}{{Lidz}
  et~al.}{2021}]{lidz2021}
{Lidz} A.,  {Chang} T.-C.,  {Mas-Ribas} L.,   {Sun} G.,  2021, \mn@doi [\apj]
  {10.3847/1538-4357/ac0af0}, \href
  {https://ui.adsabs.harvard.edu/abs/2021ApJ...917...58L} {917, 58}

\bibitem[\protect\citeauthoryear{{Mason} \& {Gronke}}{{Mason} \&
  {Gronke}}{2020}]{mason2020}
{Mason} C.~A.,  {Gronke} M.,  2020, \mn@doi [\mnras] {10.1093/mnras/staa2910},
  \href {https://ui.adsabs.harvard.edu/abs/2020MNRAS.499.1395M} {499, 1395}

\bibitem[\protect\citeauthoryear{{Mason}, {Treu}, {Dijkstra}, {Mesinger},
  {Trenti}, {Pentericci}, {de Barros}  \& {Vanzella}}{{Mason}
  et~al.}{2018}]{mason2018}
{Mason} C.~A.,  {Treu} T.,  {Dijkstra} M.,  {Mesinger} A.,  {Trenti} M.,
  {Pentericci} L.,  {de Barros} S.,   {Vanzella} E.,  2018, \mn@doi [\apj]
  {10.3847/1538-4357/aab0a7}, \href
  {https://ui.adsabs.harvard.edu/abs/2018ApJ...856....2M} {856, 2}

\bibitem[\protect\citeauthoryear{{Mertens} et~al.,}{{Mertens}
  et~al.}{2020}]{mertens2020}
{Mertens} F.~G.,  et~al., 2020, \mn@doi [\mnras] {10.1093/mnras/staa327}, \href
  {https://ui.adsabs.harvard.edu/abs/2020MNRAS.493.1662M} {493, 1662}

\bibitem[\protect\citeauthoryear{{Mesinger} \& {Furlanetto}}{{Mesinger} \&
  {Furlanetto}}{2007}]{mesinger2007}
{Mesinger} A.,  {Furlanetto} S.,  2007, \mn@doi [\apj] {10.1086/521806}, \href
  {https://ui.adsabs.harvard.edu/abs/2007ApJ...669..663M} {669, 663}

\bibitem[\protect\citeauthoryear{{Mesinger}, {Furlanetto}  \& {Cen}}{{Mesinger}
  et~al.}{2011}]{mesinger2011}
{Mesinger} A.,  {Furlanetto} S.,   {Cen} R.,  2011, {21cmFAST: A Fast,
  Semi-Numerical Simulation of the High-Redshift 21-cm Signal}, Astrophysics
  Source Code Library, record ascl:1102.023 (\mn@eprint {ascl} {1102.023})

\bibitem[\protect\citeauthoryear{{Miralda-Escud{\'e}}}{{Miralda-Escud{\'e}}}{1998}]{miralda-escude1998}
{Miralda-Escud{\'e}} J.,  1998, \mn@doi [\apj] {10.1086/305799}, \href
  {https://ui.adsabs.harvard.edu/abs/1998ApJ...501...15M} {501, 15}

\bibitem[\protect\citeauthoryear{{Planck Collaboration} et~al.,}{{Planck
  Collaboration} et~al.}{2020}]{planck2020}
{Planck Collaboration} et~al., 2020, \mn@doi [\aap]
  {10.1051/0004-6361/201833910}, \href
  {https://ui.adsabs.harvard.edu/abs/2020A&A...641A...6P} {641, A6}

\bibitem[\protect\citeauthoryear{{Rudd}, {Zentner}  \& {Kravtsov}}{{Rudd}
  et~al.}{2008}]{rudd2008}
{Rudd} D.~H.,  {Zentner} A.~R.,   {Kravtsov} A.~V.,  2008, \mn@doi [\apj]
  {10.1086/523836}, \href
  {https://ui.adsabs.harvard.edu/abs/2008ApJ...672...19R} {672, 19}

\bibitem[\protect\citeauthoryear{{Tang} et~al.,}{{Tang}
  et~al.}{2023}]{tang2023}
{Tang} M.,  et~al., 2023, \mn@doi [arXiv e-prints] {10.48550/arXiv.2301.07072},
  \href {https://ui.adsabs.harvard.edu/abs/2023arXiv230107072T} {p.
  arXiv:2301.07072}

\bibitem[\protect\citeauthoryear{{Totani}, {Aoki}, {Hattori}  \&
  {Kawai}}{{Totani} et~al.}{2016}]{totani2016}
{Totani} T.,  {Aoki} K.,  {Hattori} T.,   {Kawai} N.,  2016, \mn@doi [\pasj]
  {10.1093/pasj/psv123}, \href
  {https://ui.adsabs.harvard.edu/abs/2016PASJ...68...15T} {68, 15}

\bibitem[\protect\citeauthoryear{{Woosley}}{{Woosley}}{1993}]{woosley1993}
{Woosley} S.~E.,  1993, \mn@doi [\apj] {10.1086/172359}, \href
  {https://ui.adsabs.harvard.edu/abs/1993ApJ...405..273W} {405, 273}

\bibitem[\protect\citeauthoryear{{Zahn}, {Mesinger}, {McQuinn}, {Trac}, {Cen}
  \& {Hernquist}}{{Zahn} et~al.}{2011}]{zahn2011}
{Zahn} O.,  {Mesinger} A.,  {McQuinn} M.,  {Trac} H.,  {Cen} R.,   {Hernquist}
  L.~E.,  2011, \mn@doi [\mnras] {10.1111/j.1365-2966.2011.18439.x}, \href
  {https://ui.adsabs.harvard.edu/abs/2011MNRAS.414..727Z} {414, 727}

\bibitem[\protect\citeauthoryear{{Zahn} et~al.,}{{Zahn} et~al.}{2012}]{spt2012}
{Zahn} O.,  et~al., 2012, \mn@doi [\apj] {10.1088/0004-637X/756/1/65}, \href
  {https://ui.adsabs.harvard.edu/abs/2012ApJ...756...65Z} {756, 65}

\bibitem[\protect\citeauthoryear{{Zhu} et~al.,}{{Zhu} et~al.}{2021}]{zhu2021}
{Zhu} Y.,  et~al., 2021, \mn@doi [\apj] {10.3847/1538-4357/ac26c2}, \href
  {https://ui.adsabs.harvard.edu/abs/2021ApJ...923..223Z} {923, 223}

\makeatother
\end{thebibliography}




\appendix




\bsp	
\label{lastpage}
\end{document}